\documentclass[reprint,
superscriptaddress,
amsmath,amssymb,
 prl,
]{revtex4-1}

\usepackage{graphicx}
\usepackage{dcolumn}
\usepackage{bm}
\usepackage[utf8]{inputenc}
\usepackage[dvipsnames]{xcolor}

\begin{document}

\title{Interfacial layering in the electric double layer of ionic liquids}

\author{J. Pedro de Souza}
\affiliation{Department of Chemical Engineering, Massachusetts Institute of Technology, Cambridge, MA, USA}

\author{Zachary A. H. Goodwin}
\affiliation{Department of Physics, CDT Theory and Simulation of Materials, Imperial College of London, South Kensington Campus, London SW7 2AZ, UK}
\affiliation{Thomas Young Centre for Theory and Simulation of Materials, Imperial College London, South Kensington Campus, London SW7 2AZ, UK}

\author{Michael McEldrew}
\affiliation{Department of Chemical Engineering, Massachusetts Institute of Technology, Cambridge, MA, USA}

\author{Alexei A. Kornyshev}
\affiliation{Department of Chemistry, Imperial College of London, Molecular Science Research Hub, White City Campus, London W12 0BZ, UK}
\affiliation{Thomas Young Centre for Theory and Simulation of Materials, Imperial College London, South Kensington Campus, London SW7 2AZ, UK}

\author{Martin Z. Bazant}
\affiliation{Department of Chemical Engineering, Massachusetts Institute of Technology, Cambridge, MA, USA}
\affiliation{Department of Mathematics, Massachusetts Institute of Technology, Cambridge, MA, USA}

\date{\today}

\begin{abstract}
Ions in ionic liquids and concentrated electrolytes reside in a crowded, strongly-interacting environment, leading to the formation of discrete layers of charges at interfaces and spin-glass structure in the bulk. Here, we propose a simple theory that accurately captures the coupling between steric and electrostatic forces in ionic liquids. The theory predicts the formation of discrete layers of charge at charged interfaces. Further from the interface, or at low charges, the model outputs slowly-decaying oscillations in the charge density with a wavelength of a single ion diameter, as shown by analysis of the gradient expansion. The gradient expansion suggests a new structure for partial differential equations describing the electrostatic potential at charged interfaces. We find quantitative agreement between the theory and molecular simulations in the differential capacitance and concentration profiles. 

\end{abstract}

\maketitle

%
%

\textit{Introduction}- The spatial organization of ions in concentrated electrolytes leads to strong density and charge oscillations in the electric double layer (EDL) at charged interfaces~\cite{CR, bazant2009towards,kirkwood1936statistical}. When the concentration is beyond the dilute limit of the established Poisson-Boltzmann (PB) theory, one must account for correlation and packing effects, particularly as the Debye length approaches the size of a single ion~\cite{EMFnDS}. Methods to correct the PB equations include the hypernetted-chain equation~\cite{patey1980interaction,lozada1982application, kjellander1992double, attard1993asymptotic, kjellander1992exact,Kj_UND}, mean-spherical approximation~\cite{DC,ROT_UND}, density functional theory~\cite{KS,IIP,TDRPM,CDFT,GAV_UND, wu2011classical, henderson2011density, ma2020classical,CIA}, and dressed-ion theory\cite{kjellander1994dressed, kanduvc2010dressed}. While many methods can accurately predict EDL profiles, they often lack the simplicity and physical transparency of the PB theory which they seek to correct~\cite{EMFnDS}. 

%
%

More recently, with the rediscovery of room temperature ionic liquids (RTILs)~\cite{TW,JH_TW} and their applications to energy storage devices~\cite{NH,CR}, the task of understanding the interfacial structure in concentrated electrolytes has surged~\cite{ILCG}. Describing the EDL of RTILs is particularly difficult because of the competition between strong steric and electrostatic forces~\cite{CR}, as illustrated in Fig. 1, and the fact that the expected Debye screening length is unphysically smaller than the diameter of an ion. In fact, the coupling of density and charge has been described as the ground state for a spin-glass Hamiltonian for ionic nearest neighbors (given their positions)~\cite{levy2019spin}, which is extremely difficult to describe with continuum equations. The interplay between ion position and charge order gives rise to the well known crossover from the overscreening regime (where decaying oscillations of charge density occur) to the crowding regime (where dense layers of countercharge accumulate at the interface before an overscreening tail)~\cite{SSS,NCS, OS, kirchner2013electrical}.

\begin{figure}[b!]
\centering
\includegraphics[width=0.9 \linewidth]{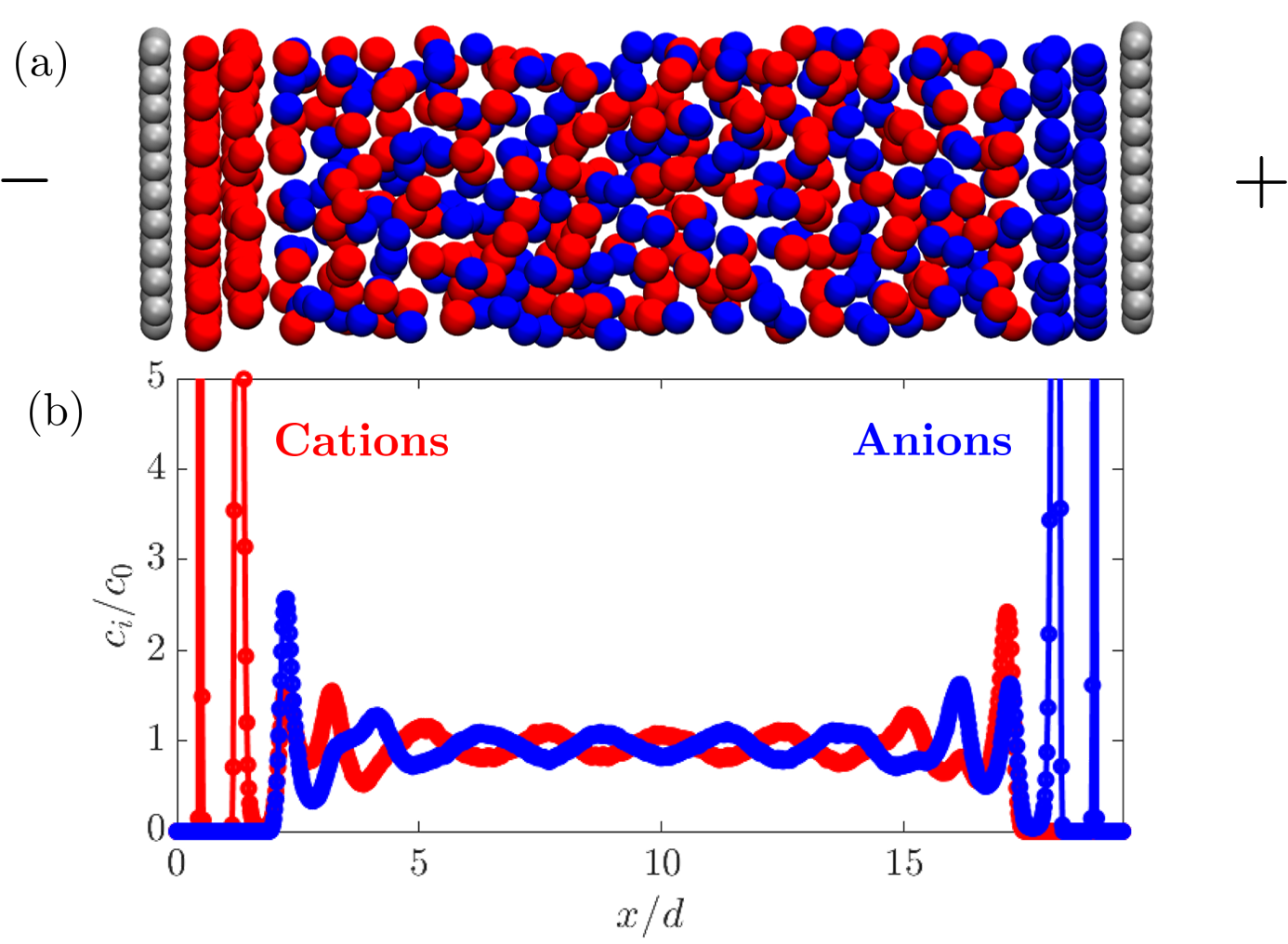}
\caption{(a) Illustration of a concentrated, crowded electrolyte forming structured double layers at high surface charge density. The cations are red, the anions are blue, and the surface atoms are shown in gray, with negative charge on the left surface and positive charge on the right surface. (b) Corresponding concentration profile for a representative room temperature ionic liquid of equal-sized hard spheres ($c_0=5$ M, $d=0.5$~nm, $\epsilon_r=10$, $q_s=120$~$\mu$C/cm$^2$, $T=300$~K).}
\end{figure}

Perhaps one of the most popular descriptions of the overscreening versus crowding problem~\cite{SSS,NCS} in RTILs is the Bazant-Storey-Kornyshev (BSK) theory~\cite{OS}. There, a higher order gradient term in electrostatic potential was proposed, in addition to the commonly used lattice-gas excluded-volume excess chemical potential~\cite{PC,SEIL}. The BSK theory has been shown to accurately describe electrostatic correlations for dilute electrolytes and counter-ion only systems~\cite{de2019continuum,misra2019theory}. \color{black} In the concentrated limit of RTILs, however, the BSK theory has some notable limitations: The screening is always short-ranged; the period of the oscillation is not necessarily the size of an ion; the number and extent of oscillations is significantly underestimated; and the formation of discrete charged layers at the interface is not captured. \color{black} More recent work has suggested that the overscreening structure is a similar concept to the finite-size~\cite{adar2019screening} and orientation of ionic aggregates~\cite{avni2020charge} near charged interfaces.

\begin{figure}[b]
\centering
\includegraphics[width=0.9 \linewidth]{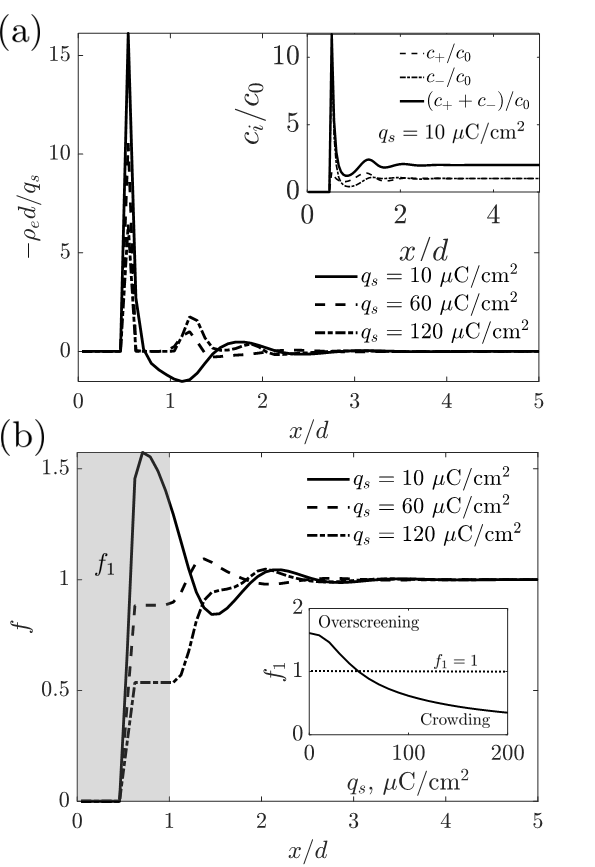}
\caption{Layering of ions in a concentrated electrolyte or ionic liquid. (a) The overscreening `signature:' the charge density of ions near a positively charged electrode scaled to the surface charge density on the electrode. The inset shows the concentration profile for each ion at $q_s=10$ $\mu$C/cm$^2$, with oscillations in both the sum of concentrations and in the difference in concentrations. (b) The cumulative charge density as a function of the distance from the interface, with inset showing the extent of screening in the first layer of charge, $f_1$. Overscreening occurs when the net charge in the first layer is larger than the charge on the electrode.}
\end{figure}

%
%
\color{black}
In this letter, we propose a free energy functional to describe the coupling between steric and electrostatic forces, and therefore, capture the ``spin-glass" nature of charge-mass correlations in RTILs. The theory predicts discrete layering, extended overscreening with a longer screening length than the size of an ion with an oscillation period of one ion diameter, and quantitative agreement with simulated differential capacitance. Our free energy functional is a new hybrid approach using the weighted density approximation to describe the finite size of ions in both their electrostatic and steric interactions. Without fitting parameters, the theory has strong predictive capabilities, and it has a similar simplicity to the other modified-Poisson-Boltzmann approaches. While we explore the equilibrium properties at interfaces, the presented formulation could be extended to RTILs out of equilibrium, phase field crystal models, or systems including a structured solvent. \color{black} 

\textit{Theory}- We modify the electrostatic and hard sphere packing free energies by representing them in terms of weighted densities of local concentrations, similar to weighted-density approximations including fundamental measure theory~\cite{roth2010fundamental, rosenfeld1989free, tarazona1985free}. We rationalize these choices by treating the ions as hard, conducting, charged spheres of finite size, with point potential:
\begin{equation}\label{eq:eqGreens}
G_i(r)=\begin{cases}\frac{z_i e}{4\pi \epsilon r} & r \geq R\\
\phi_0 & r<R
\end{cases}
\end{equation}
where $\phi_0$ is a constant within a given ion, $\epsilon$ is the permittivity surrounding the ion \color{black}(assumed constant in this work as an average effective background value), \color{black} $z_i e$ is the charge of the ion, $R$ is the radius of an ion, and $r$ is the distance from the center of an ion. \color{black} The physical basis for the Ashcroft pseudopotential character~\cite{ashcroft1966electron} of the Green's function is that the electrostatic potential within a finite-sized ion is effectively overwhelmed by the hard sphere potential within the ion. Therefore, the electrostatic potential is an undefined constant within the sphere and can decay as a $1/r$ potential only beyond the ionic radius. \color{black} The linear integro-differential equation corresponding to this Green's function is:
\begin{equation} 
\label{eq:eqMPB}
\begin{split}
    &\epsilon\nabla^2 \phi=-\bar{\rho}_e(\mathbf{r})=-\int d\mathbf{r^\prime} \rho_e(\mathbf{r}) w_s(\mathbf{r}-\mathbf{r^\prime})\\
    & w_s(\mathbf{r}-\mathbf{r^\prime})=\frac{1}{4\pi R^2}\delta\left(R-\mid\mathbf{r}-\mathbf{r^\prime}\mid\right)
\end{split}
\end{equation}
which is the key modified \color{black} mean-field \color{black} Poisson equation in our work. Here $\phi$ is the electrostatic potential, $\rho_e=\sum_i z_i e c_i$ is the charge density of ionic centers, \color{black}$c_i$ is the number density of the centers of species $i$, \color{black} $\bar{\rho}_e$ is the weighted charge density (charge density calculated for the smeared charge of an ion over its surface), and $w_s$ is the weighting function. Integrating contributions of the smeared charges results in the ``actual" charge density which resides in the Poisson equation. While our weight function for the charge density resembles the choice of charge form factor in Ref.~\citenum{adar2019screening} for ionic screening in the bulk, we construct a mean-field equation that gives the ionic density at a flat interface at high charge density.

From the above modified Poisson equation, the electrostatic free energy density becomes:
\begin{equation}
    \mathcal{F}^\mathrm{el}[\bar{\rho}_e, \phi] =  \int d\mathbf{r} \Big\{-\frac{\epsilon}{2}(\nabla\phi)^2+ \bar{\rho}_e \phi\Big\}.
\end{equation}
The chemical part of the free energy contains an ideal \color{black}entropic \color{black} contribution: $\mathcal{F}^{\mathrm{id}}[\{c_i(\mathbf{r})\}]=\sum_i  k_{B}T \int d\mathbf{r}\,c_i(\mathbf{r})\big[\ln (\Lambda^3 c_i(\mathbf{r})) - 1\big]$, where $k_BT$ is thermal energy and $\Lambda$ is the thermal de Broglie wavelength \cite{roth2010fundamental}. There is also an excess contribution from crowding of the finite-sized ions. The Carnahan-Starling equation of state accurately describes the properties of hard sphere liquids. Here, we adapt it and assume that the local excess free energy depends on volumetrically weighted densities, similar to fundamental measure theory\cite{roth2010fundamental,rosenfeld1989free}:
\begin{equation}
\mathcal{F}^\mathrm{ex}[\bar{c}_i(\mathbf{r})]=\frac{k_{B}T}{v}\int d\mathbf{r}\,\bigg[\dfrac{1}{1 - \bar{p}} - 3\bar{p} + \dfrac{1}{(1 - \bar{p})^{2}}\bigg]
\end{equation}
where $\bar{p}=\sum_i v \bar{c}_i$ is the weighted volumetric filling fraction and $v=4\pi R^3/3$ the volume of an ion. The weighted densities are defined by:
\begin{equation}
    \begin{split}
        &\bar{c}_i(\mathbf{r})=\int d\mathbf{r^\prime} c_i(\mathbf{r}) w_v(\mathbf{r}-\mathbf{r^\prime})\\ &w_v(\mathbf{r}-\mathbf{r}^\prime)=\frac{1}{v}\Theta\left(R-\mid\mathbf{r}-\mathbf{r^\prime}\mid\right)
    \end{split}
\end{equation}
where the scalar valued weighting function has units of inverse volume, \color{black} and the function $\Theta$ represents a Heaviside step function. \color{black} Therefore, the densities with which the mean field electrostatic interaction or hard sphere interaction occurs are computed with a quantized volume of one ion.     
\color{black} Physically, the free energy is infinite as the volumetric-weighted filling fraction goes to one~\footnote{The weighted density is necessary to describe the formation of discrete layers of charge at high surface charge density, which cannot be captured by local-density approximations.}. \color{black} For the purposes of this study, the electrostatic weighting function will be homogenized on a surface of an ionic sphere, whereas the volumetric packing fraction will be homogenized over a volume of an ionic sphere.

Minimizing the free energy functional, we arrive at a modified PB equation, Eq.~\eqref{eq:eqMPB}, where the distribution of ion (center) densities are determined by
\begin{equation}\label{eq:eqBoltz}
    c_i=c_{i,0}\exp(-z_i\beta e\bar{\phi}-\beta\bar{\mu}^\mathrm{ex}_i+\beta{\mu}^\mathrm{ex}_{i,bulk})
\end{equation}
with $\beta$ as the inverse thermal energy, $\bar{\phi}=\phi*w_s$ and $\bar{\mu}^\mathrm{ex}_i={\mu}^\mathrm{ex}_i*w_v$, \color{black}(with $*$ denoting convolution), \color{black}  and excess chemical potential defined as $\beta\mu^\mathrm{ex}_i= (8\bar{p} - 9\bar{p}^{2} + 3\bar{p}^{3})/(1 - \bar{p})^{3}$ \footnote{A continuum theory of this kind does not require distinguishing ‘free,' ‘paired,' or ‘clustered’ ions \cite{mceldrew2020theory}. The ionic associations are reflected in oscillating charge density distributions.}.

\begin{figure}[b]
\centering
\includegraphics[width=0.9 \linewidth]{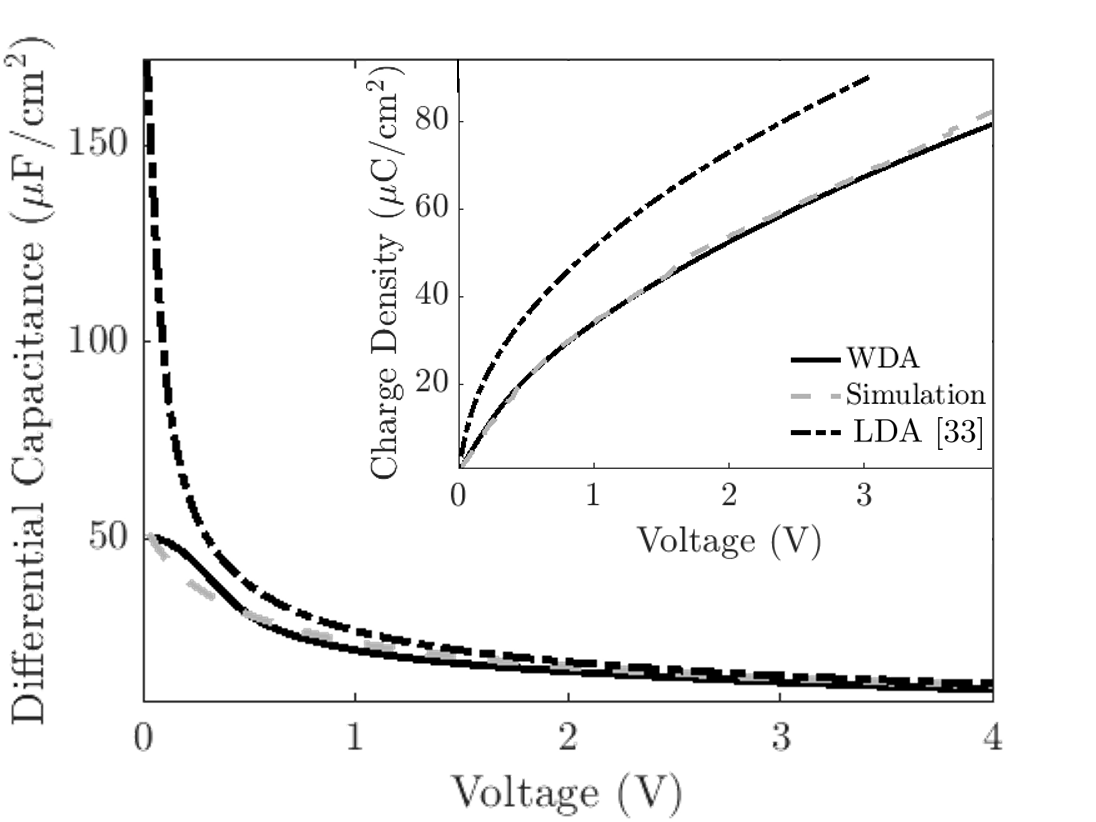}
\caption{Differential capacitance of the EDL as a function of the applied voltage, for the weighted density approximation (WDA) in Eq.~\eqref{eq:eqMPB}, simulations, and the \color{black}local density approximation (LDA) \color{black} formula~\cite{PC}, given in the SM~\cite{SM_ref}. \textit{Inset:} The charge density in the double layer as a function of the applied voltage. The parameters are identical to Fig. 2.}
\end{figure}

\textit{Results and Discussion}- We solve the above coupled integro-differential equations \ref{eq:eqMPB} and \ref{eq:eqBoltz} at a flat electrode, with surface charge density, $q_s$, at $x=0$. In this case, the standard boundary condition for the potential is applied $\mathbf{\hat{n}}\cdot\epsilon\nabla\phi\rvert_s=-q_s$. The local ionic densities (of centers) $c_i$ and charge density (of ionic centers) $\rho_e$ are assumed to be zero within one radius from the surface, from $x=0$ to $x=R$, due to hard sphere exclusion. We solve for the area averaged density, and we therefore reduce all equations to be dependent on one coordinate, $x$. Numerically, we discretize the equations using a simple finite difference approach, similar to how the standard PB equations could be solved. More details on the numerics are provided in the Supplemental Material (SM)~\cite{SM_ref}. 

For further intuition, we analyze a simple gradient expansion of the weighting functions that turns them into operators: $w_j=1+{\ell_j}^2\nabla^2$, where $\ell_j$ is given by $\ell_s=d/\sqrt{24}$ for $w_s$ and $\ell_v=d/\sqrt{40}$ for $w_v$, as derived in the SM~\cite{SM_ref}. The corresponding free energy density is given by:
\begin{equation}
    \mathcal{F}^\mathrm{el}[\bar{\rho}_e, \phi] =  \int d\mathbf{r} \Big\{-\frac{\epsilon}{2}(\nabla\phi)^2+ \rho_e \phi-{\ell_s}^2\nabla\rho_e\cdot \nabla \phi \Big\}.
\end{equation} 
The leading order term in the expansion corresponds to a dipole density interacting with an electric field, interpretable as ionic pairs of effective volumetric dipole moment ${\ell_s}^2\nabla\rho_e$, \color{black} an effective polarization vector formed by gradients in the local charge density $\rho_e$~\cite{avni2020charge}. \color{black}Note that since the order of the differential equation increases, we need an additional boundary condition. We assume this to be $\mathbf{n}\cdot \nabla \rho_e\rvert_s=0$ in order to satisfy electroneutrality in the differential equation, namely that: $\int d\mathbf{r} \rho_e(\mathbf{r})=-\int d\mathbf{r_s} q_s (\mathbf{r_s})$.

\color{black}The above gradient expansion does not reproduce the profile at the initial contact of the ionic liquid with the surface. In particular, the differential form cannot capture the discontinuous contact point at $x=R$, and so the solutions are shifted by one ionic radius. Even so, the gradient expansion is valid farther from the surface and is useful for deriving analytical approximations for the theory. \color{black} Furthermore, the differential form may be easier to apply to problems in diverse applications such as electrokinetics~\cite{storey2012effects}, colloidal interactions~\cite{misra2019theory}, or electrochemical storage~\cite{zhao2011diffuse, mceldrew2018theory} than the full integro-differential theory~\cite{TUE}.  As an example, we will first analyze the gradient expansion of the continuum theory in terms of its limiting linear response behavior, which asymptotically matches the behavior of the full integral equation far from the interface. Further comparisons are included in the SM~\cite{SM_ref}.

\begin{figure}[t]
\centering
\includegraphics[width=0.85 \linewidth]{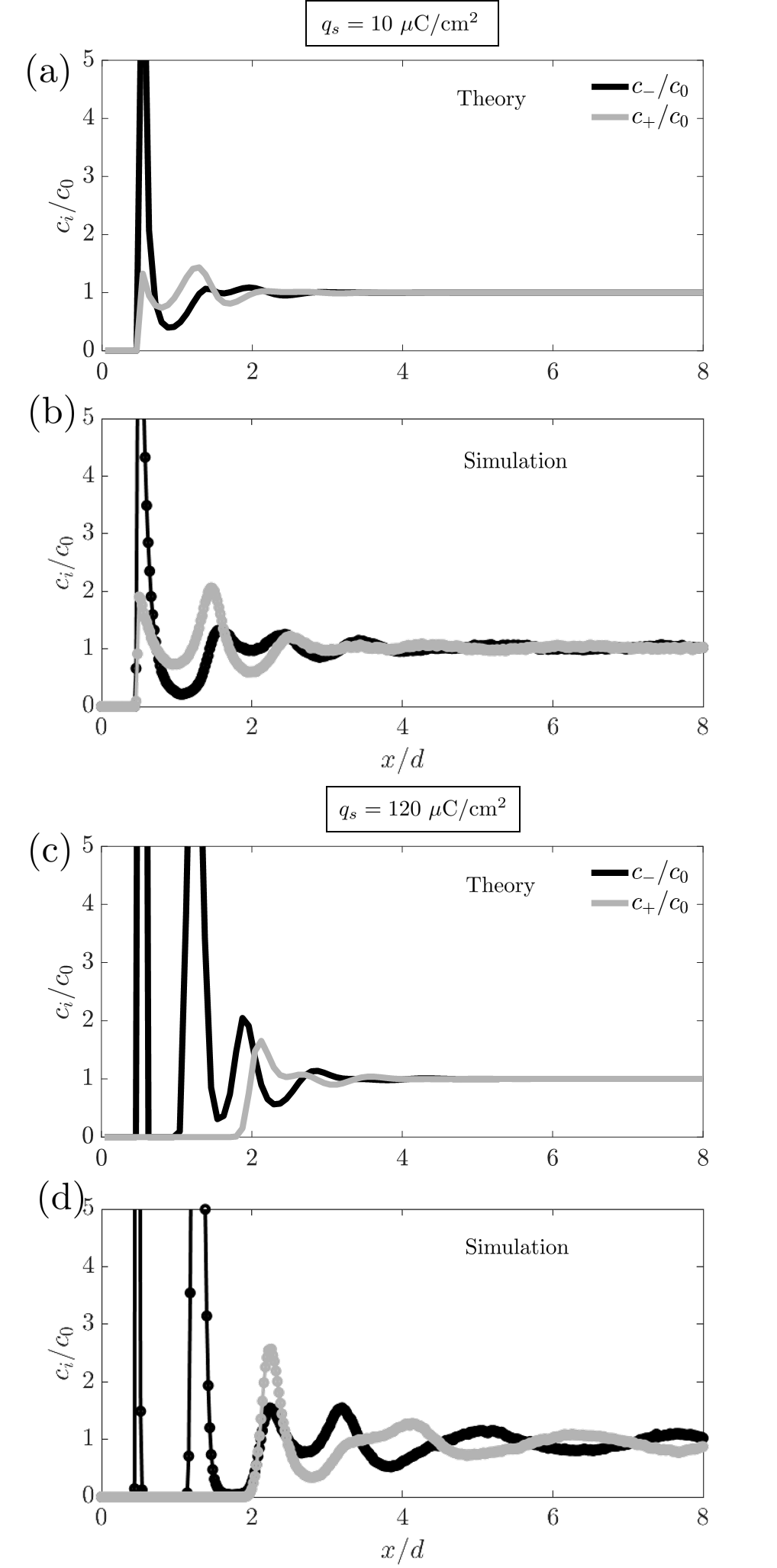}
\caption{Comparison of theory (a,c) and simulation (b,d) concentration profiles for two different charge densities: $q_s=10$~$\mu$C/cm$^2$ and $q_s=120$~$\mu$C/cm$^2$. The electrolyte has the same parameters as in Figs. 2 and 3.}
\end{figure}

In linear response, the equation for the potential is:
\begin{equation} 
    \lambda_D^2\nabla^2\phi-(1+\ell_s^2\nabla^2)^2\phi=0.
    \label{eq:eqlin}
\end{equation}
\color{black} where $\lambda_D$ is the Debye length.\color{black} While the equation is fourth order, similar to the linearized BSK equation, it has different decaying modes due to an additional second order term. The eigenvalues of the above differential equation, \color{black}denoted by the inverse decay length $\kappa_s=1/\lambda_s$, \color{black} have the form:
\begin{equation}
    \kappa_s \lambda_D=\frac{1\pm\sqrt{1-4(\ell_s/\lambda_D)^2}}{2(\ell_s/\lambda_D)^2}.
\end{equation}
Note that the form of Eq.~\eqref{eq:eqlin} bears some resemblance to the Swift-Hohenberg equation~\cite{swift1977hydrodynamic}, commonly used to describe pattern formation and other phase-field crystal models~\cite{elder2004modeling}; here electrostatics and finite size drive the pattern formation. When $\ell_s/\lambda_D>1/2$, oscillations appear in the solution, and in the limit of $\ell_s/\lambda_D\gg1/2$, the screening length takes the form: $ \kappa_s \lambda_D={\lambda_D^2}/{\ell_s^2}\pm i {\lambda_D}/{\ell_s}$. At high concentration, the ions will therefore form charge density layers on the scale of the ionic size, with period of 1.28 $d$, similar to the result from simulations. In strongly correlated regimes, the real part of the screening length will scale as: $ \ln\left[\mathrm{Re}\left({\lambda_s}/{\lambda_D}\right)\right]=2\ln\left({d}/{\lambda_D}\right)+\mathrm{const}$, increasing with concentration. This result is qualitatively in agreement with surface force experiments~\cite{UND,NMD}, but they find a scaling factor 3 rather than 2. They also measure monotonic decay, and not decaying oscillations in the overscreening tail as predicted by the theory. Note that the mass density oscillations also have a characteristic decay length, but it is decoupled from the electrostatic potential at linear response for ions of the same size, as discussed in the SM~\cite{SM_ref}. The discrepancy in exponents may be due to the symmetric size of ions in the analysis here, which limits the coupling.

Next, we compute the ion concentration and density profiles as a function of charge density for some model parameters ($c_0=5$~M,  $d=0.5$~nm, $\epsilon_r=10$, $T=300$~K), shown in Fig. 2. Note the parameters shown here are meant to be representative of RTILs, but the simplifying assumptions of similarly-sized cations and anions prevent a direct comparison with experimental results for asymmetric ionic liquids~\footnote{The symmetric size limits the possible parameter space due to maximal packing constraints, leading to a high differential capacitance at zero charge for both the theory and the simulation relative to experiments on real RTILs.}. We also present the cumulative screening charge, defined as $f(x)=-\int_0^x \rho_e (x^\prime)dx^\prime/q_s$. At low surface charge density, the first layer of charge has about 57\% more counter charge than the surface charge. Subsequent layers of alternating charge are formed. At low surface charge density, the ion concentrations themselves are affected by overall structuring of the fluid ($c_++c_-$) due to packing at the interface. At higher charge density, the inhibitive force of packing at the interface decreases the extent of overscreening in the first layer, $f_1$. Eventually, as the charge density exceeds the total amount of charge that can be stored in a single layer of ions, a secondary layer is formed. When this occurs, the extent of overscreening becomes determined by the renormalized charge on the interface. The chosen simulation parameters are in the strongly oscillating regime $\ell_s/\lambda_D\approx 2.1$, meaning that the far range screening tail has approximate wavelength of one ionic diameter and long decay length.

It is instructive to compare the predictions of the theory to MD simulations of a Lennard-Jones electrolyte with the same parameters. The differential capacitance, $C= \mid d q_s/d\phi_0\mid$ is evaluated in Fig. 3 as a function of the potential at $x=0$, $\phi_0$. Compared to simulations, the weighted density theory captures the low capacitance at zero charge and the decay of capacitance at large voltages. The theory presented here agrees much better with simulations compared to the \color{black} local density approximation \color{black} formula~\cite{PC,EA}; the improvements in the crowding regime, at large voltages, are due to use of the weighted Carnahan-Starling approximation rather than the simple \color{black} local density approximation \color{black} formula, both obeying, however, the $V^{-1/2}$ limiting law~\cite{PC, TUE}. In Fig. 4, the layering structure is compared between theory and simulation for low and high charge densities. The theory is able to qualitatively match the structuring in the simulations, with charge density oscillations and eventually layers of the same charge at high charge density. Even so, the wavelength in the charge density oscillations are off by about a factor of 1.3. \color{black}Such a discrepancy could be captured by changing the form of $w_s$ to extend beyond the size of the ionic radius, but modifications to $w_s$ are not considered in this work~\footnote{A different form of $w_s$ implies different physical assumptions for the Green's function in equation \ref{eq:eqGreens}}. \color{black}


The developed continuum theory captures the key points in the interplay between overscreening and crowding in EDL of ionic liquids, including: 1) Decaying charge density profiles near the electrode and the overscreening effect as a consequence of molecular layering, 2) The onset of crowding through the shift of the overscreening to a third, and then subsequently further layers, and 3) The emergence of the long range screening tail in ultraconcentrated ionic systems \footnote{The integral theory also performs better than the BSK theory \cite{OS} in terms of describing the layered structure at the interface.}.
 


\textit{Acknowledgements}- All authors acknowledge the support from the MIT-Imperial College Seed Fund. JPD acknowledges support from the Center for Enhanced Nanofluidic Transport, an Energy Frontier Research Center funded by the U.S. Department of Energy, Office of Science, Basic Energy Sciences under Award \# DE-SC0019112 (supporting simulations), and from the National Science Foundation Graduate Research Fellowship under award number \#1122374 (supporting theory development). MM was supported by the Amar G. Bose Research Grant. ZAHG was supported through a studentship in the Centre for Doctoral Training on Theory and Simulation of Materials at Imperial College London funded by the EPSRC (EP/L015579/1) and from the Thomas Young Centre under grant number TYC-101.

\bibliography{REF.bib}

\end{document}


\title{Supplemental Material: Interfacial layering in the electric double layer of ionic liquids}

\author{J. Pedro de Souza}
\affiliation{Department of Chemical Engineering, Massachusetts Institute of Technology, Cambridge, MA, USA}

\author{Zachary A. H. Goodwin}
\affiliation{Department of Physics, CDT Theory and Simulation of Materials, Imperial College of London, South Kensington Campus, London SW7 2AZ, UK}
\affiliation{Thomas Young Centre for Theory and Simulation of Materials, Imperial College London, South Kensington Campus, London SW7 2AZ, UK}

\author{Michael McEldrew}
\affiliation{Department of Chemical Engineering, Massachusetts Institute of Technology, Cambridge, MA, USA}

\author{Alexei A. Kornyshev}
\affiliation{Department of Chemistry, Imperial College of London, Molecular Science Research Hub, White City Campus, London W12 0BZ, UK}
\affiliation{Thomas Young Centre for Theory and Simulation of Materials, Imperial College London, South Kensington Campus, London SW7 2AZ, UK}

\author{Martin Z. Bazant}
\affiliation{Department of Chemical Engineering, Massachusetts Institute of Technology, Cambridge, MA, USA}
\affiliation{Department of Mathematics, Massachusetts Institute of Technology, Cambridge, MA, USA}

\date{\today}

\renewcommand{\theequation}{S\arabic{equation}}
\renewcommand{\thefigure}{S\arabic{figure}}

\maketitle

\section{Local and weighted density profiles}

\begin{figure*}[h]
\centering
\includegraphics[width=1 \linewidth]{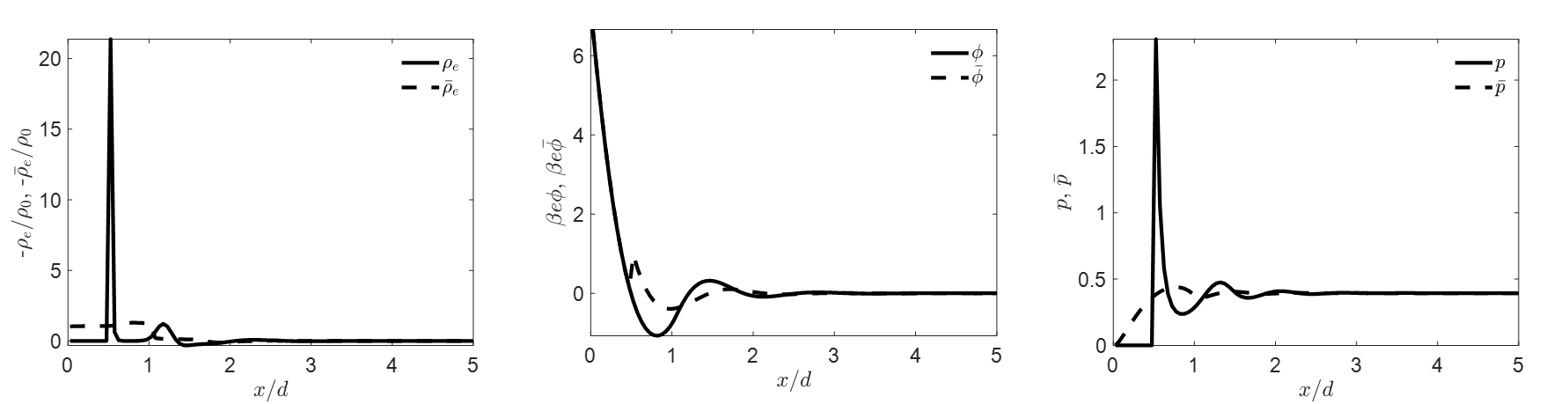}
\caption{Comparison of local (\solidrule) and weighted variables (\dashedrule) for (a) charge density (b) electrostatic potential, and (c) filling fraction. The parameters are identical to Fig. 2, and the surface charge density is fixed for (a-c) at 60 $\mu$C/cm$^2$.}
\label{fig:S1}
\end{figure*}

The key feature of the theory is the convolution of the charge density, electrostatic potential, and filling fraction with their corresponding weighting functions:
\begin{align}
    &\bar{\rho}_e=\int d\mathbf{r^\prime} \rho_e(\mathbf{r}) w_s(\mathbf{r}-\mathbf{r^\prime})\\
    &\bar{\phi}=\int d\mathbf{r^\prime} \phi(\mathbf{r}) w_s(\mathbf{r}-\mathbf{r^\prime})\\
    &\bar{p}=\int d\mathbf{r^\prime} p(\mathbf{r}) w_v(\mathbf{r}-\mathbf{r^\prime})
\end{align}
In Fig.~\ref{fig:S1}, each of the weighted and local variables are compared. Evidently, the sharpness apparent in the local profiles is smoothed out by the weighting function. In other words, the weighted densities allow sharp layers to form at the interface without breaking the constraint of maximal packing or overriding the strong electrostatic force towards electroneutrality.  

\section{Variation of parameters}

\begin{figure*}[h]
\centering
\includegraphics[width=1 \linewidth]{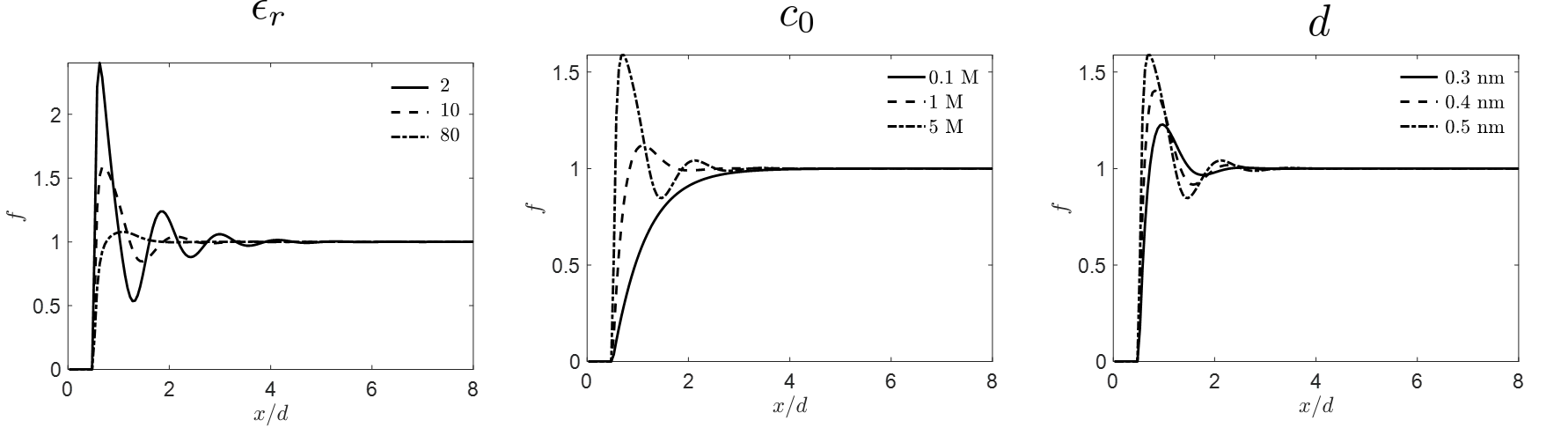}
\caption{The effect of variation of parameters in theory. The cumulative charge is shown versus the distance from the charged interface, for low surface charge densities ($q_s\rightarrow 0$). At high relative permittivity, low concentration, or small diameters, the PB theory is recovered with no oscillations. In the opposite limits, more oscillations of charge are predicted by the theory. The parameters are identical to Fig. 2 in the main text, with the exception of the one being varied.}
\label{fig:S2}
\end{figure*}

The simple modified Poisson equation posed in the main text has a rich solution structure that incorporates the essential length scales in the problem, including the ion diameter. Here, we vary some of the main parameters in the theory to observe what electrical double layer structure is expected in the different regimes, as shown in Fig.~\ref{fig:S2}. Namely we vary the relative permittivity, the bulk concentration, and the ion diameter. As electrostatic forces increase (corresponding to a decrease in $\epsilon_r$) the solution becomes more oscillatory, and overscreening is increased. Similarly, as concentration increases, more oscillations are expected. Finally, as the diameter increases, the layered structure becomes more prominent. Notice that the wavelength of the layers quickly converges towards the size of a single ion. Also, the non-oscillatory profiles of the PB theory are recovered as $\epsilon_r\rightarrow \infty$, $c_0\rightarrow 0$, and $d\rightarrow 0$, when the Debye length is much larger than the ion diameter.

\section{Comparison of differential form to integro-differential form}

The integral equation written here can be reduced to a differential form by retaining only the first two terms in the expanded weighting function. \begin{equation}
    w_j\approx 1+{\ell_j}^2\nabla^2
\end{equation}
\color{black}The value of $\ell_j$ is determined by the expansion of the weighting functions in Fourier space:
\begin{equation}
\begin{split}
        &w_s(k)=\frac{\sin(kR)}{kR}=1-\frac{R^2}{6} k^2+\dots\\
        &w_v(k)=\frac{3\sin(kR)-3 kR \cos(kR)}{(kR)^3}=1-\frac{R^2}{10} k^2 +\dots 
\end{split}
\end{equation}
where the above expressions match those presented in \cite{rosenfeld1989free}. Based on the above expansions, we attain:
\begin{equation}
    \begin{split}
        &\ell_s=d/\sqrt{24} \\
        &\ell_v=d/\sqrt{40}
    \end{split}
\end{equation}

If we now plug in the differential form into the electrostatic free energy from equation 3 in the main text, we get the following:
\begin{equation}
    \mathcal{F}^\mathrm{el}[\bar{\rho}_e, \phi] =  \int d\mathbf{r} \Big\{-\frac{\epsilon}{2}(\nabla\phi)^2+ \rho_e \phi+\ell_s^2\phi\nabla^2\rho_e\Big\}.
\end{equation}
On the final term, we can use the vector identity:
\begin{align}
    \nabla \cdot(b\mathbf{u})=b\nabla\cdot \mathbf{u}+\mathbf{u} \cdot \nabla b
\end{align}
where b is a scalar ($\phi$) and $\mathbf{u}$ is a vector ($\nabla \rho_e$). Making a subsitution, the free energy becomes:
\begin{equation}
    \mathcal{F}^\mathrm{el}[\bar{\rho}_e, \phi] =  \int d\mathbf{r} \Big\{-\frac{\epsilon}{2}(\nabla\phi)^2+ \rho_e \phi-\ell_s^2 \nabla \rho_e\cdot\nabla\phi+\nabla\cdot\left(\ell_s^2\phi\nabla\rho_e\right)\Big\}.
\end{equation}
Invoking the divergence theorem and the boundary condition of $\mathbf{n}\cdot \nabla \rho_e\rvert_s=0$, the surface term goes to zero, leaving the equation 7 from the main text.
\begin{equation}
    \mathcal{F}^\mathrm{el}[\bar{\rho}_e, \phi] =  \int d\mathbf{r} \Big\{-\frac{\epsilon}{2}(\nabla\phi)^2+ \rho_e \phi-\ell_s^2 \nabla \rho_e\cdot\nabla\phi\Big\}.
\end{equation}
\color{black}Taking the variation with respect to the potential and the ion concentrations, the approximate modified PB equation is:
\begin{equation}
    \begin{split}
         &\epsilon\nabla^2 \phi=-\rho_e-{\ell_s}^2\nabla^2\rho_e\\
         &c_i=c_{i,0}\exp\left(-z_i\beta e\phi-z_i\beta e {\ell_s}^2\nabla^2\phi-\beta \mu_i^\mathrm{ex}+\beta\mu_{i,bulk}^\mathrm{ex}-\beta {\ell_v}^2\nabla^2\mu_i^\mathrm{ex}\right)\\
         &\beta\mu_i^\mathrm{ex}=\frac{ 8\bar{p} - 9\bar{p}^{2} + 3\bar{p}^{3}}{(1 - \bar{p})^{3}}\\
         &\bar{p}=v\sum_i c_i +{\ell_v}^2\nabla^2c_i
    \end{split}
\end{equation}
The expansion of the weighting function is only valid when $d\ll \lambda_D$. For small deviations from the bulk density, the set of equations can be linearized, giving two decoupled differential equations for the potential $\phi$ and the mass density $p$:
\begin{equation}
\begin{split}
    &{\lambda_D}^2\nabla^2\phi=(1+{\ell_s}^2\nabla^2)^2\phi\\
    &\left[1+\frac{2\eta(4-\eta)}{(1-\eta)^4}(1+{\ell_v}^2\nabla^2)^2\right](p-\eta)=0
    \end{split}
\end{equation}

Note that the mass density and potential equations would be couple together if the ionic sizes were asymmetric, or due to any asymmetry in their excess chemical potential, potentially leading to longer screening lengths where mass density oscillations propagate charge density oscillations, and vice-versa.

Here, we analyze the equation for the potential at low surface potential, as presented in the main text, then we comment on the decay of mass density oscillations at the end of this section.
\begin{equation}
    {\lambda_D}^2\nabla^2\phi=(1+{\ell_s}^2\nabla^2)^2\phi
\end{equation}

Applying boundary conditions at $d/2$:
\begin{equation}
    \frac{d\phi}{dx}\Bigg\rvert_{x=\frac{d}{2}}=\frac{q_s}{\epsilon},\quad\quad \frac{d\phi}{dx}\Bigg\rvert_{x=\frac{d}{2}}+{\ell_s}^2\frac{d^3\phi}{dx^3}\Bigg\rvert_{x=\frac{d}{2}}=0
\end{equation}
and at $x\rightarrow \infty$:
\begin{equation}
    \quad\quad\quad \frac{d\phi}{dx}\Bigg\rvert_\infty=\frac{d^3\phi}{dx^3}\Bigg\rvert_\infty=0,
\end{equation}
an analytical formula can be derived for the potential at a charged interface for $x>d/2$:
\begin{equation}
    \phi=\phi_{d/2}\left[A_1\exp\left(-\kappa_1 (x-d/2)\right)+A_2\exp(-\kappa_2 (x-d/2))\right]
\end{equation}
for $\ell_s/\lambda_D<0.5$ and
\begin{equation}
    \phi=\phi_{d/2}\exp(-\kappa_3 (x-d/2))\left[\cos(\kappa_4 (x-d/2))+B\sin(\kappa_4 (x-d/2))\right]
\end{equation}
for $\ell_s/\lambda_D>0.5$, where the constants $\kappa_1$, $\kappa_2$, $\kappa_3$, $\kappa_4$, $A_1$, $A_2$, and $B$ are given by:
\begin{align}
    &\kappa_1 \lambda_D=\frac{1+\sqrt{1-4(\ell_s/\lambda_D)^2}}{2(\ell_s/\lambda_D)^2}\\
    &\kappa_2 \lambda_D=\frac{1-\sqrt{1-4(\ell_s/\lambda_D)^2}}{2(\ell_s/\lambda_D)^2}\\
    &\kappa_3\lambda_D=\frac{1}{2(\ell_s/\lambda_D)^2} \\
    &\kappa_4\lambda_D=\frac{\sqrt{1-4(\ell_s/\lambda_D)^2}}{2(\ell_s/\lambda_D)^2}\\
    & A_1= \frac{\kappa_2+{\ell_s}^2{\kappa_2}^3}{\kappa_2-\kappa_1+{\ell_s}^2\left({\kappa_2}^3-{\kappa_1}^3\right)}\\
    & A_2=\frac{\kappa_1+{\ell_s}^2{\kappa_1}^3}{\kappa_1-\kappa_2+{\ell_s}^2\left({\kappa_1}^3-{\kappa_2}^3\right)}\\
    & B=\frac{\kappa_3-3{\ell_s}^2{\kappa_4}^2\kappa_3+{\ell_s}^2{\kappa_3}^3}{\kappa_4+3{\ell_s}^2{\kappa_3}^2\kappa_4-{\ell_s}^2{\kappa_4}^3}
\end{align}

Interestingly, the capacitance in the diffuse part of the double layer ($x>d/2$) corresponds to the same value as the Debye capacitance:
\begin{equation}
    C_D=\frac{d q_s}{d\phi_{d/2}}=\frac{\epsilon}{\lambda_D}
\end{equation}
where in this case, the surface potential is evaluated at $x=d/2$. Note that additional capacitance would appear due to the potential drop within the distance of closest approach of the ions, giving some Stern capacitance in series with the diffuse capacitance.

\begin{figure*}[h]
\centering
\includegraphics[width=1 \linewidth]{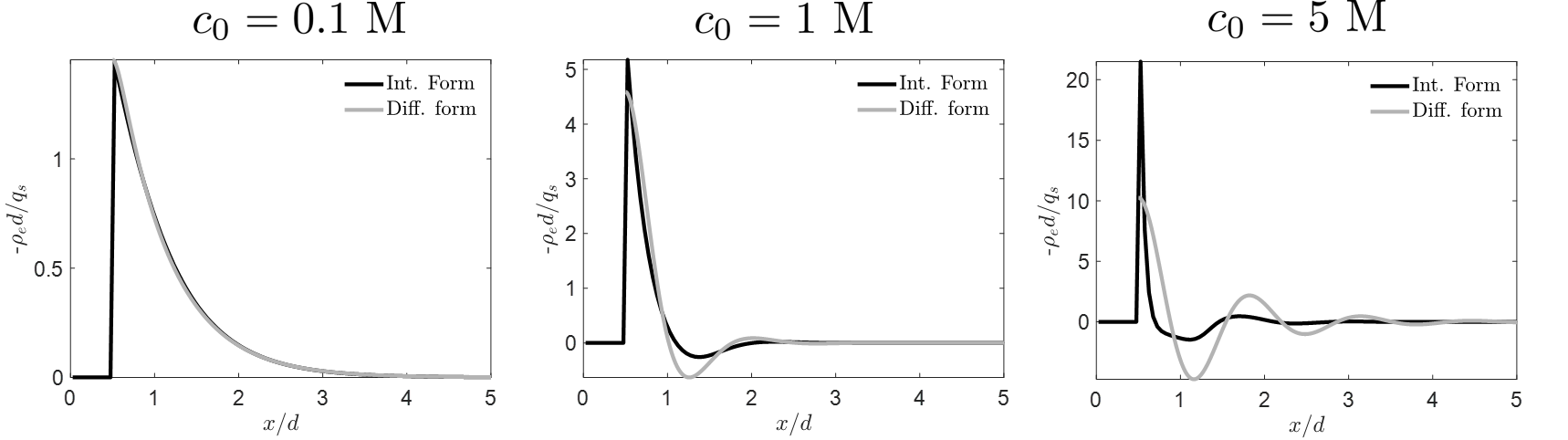}
\caption{Comparison of integral and differential form for (a) 0.1 M (b) 1 M, and (c) 5 M. The differential form performs worse at as $\ell_s/\lambda_D$ increases as concentration increases. }
\label{fig:S3}
\end{figure*}

In Fig.~\ref{fig:S2}, the profile is compared as $q_s\rightarrow 0$. The parameters are  $d=0.5$ nm, $\epsilon_r=10$, and $T=300 $ K for a 1:1 electrolyte, while the concentration of the ions are varied. As the system becomes more correlated, the predictions of the gradient expanded theory become less accurate, as shown in Fig.~\ref{fig:S3}. Even so, the gradient expanded theory should describe the decay of the potential and profiles far from the interface. In the strong coupling limit, the predictions for the capacitance should also diverge from the gradient expansion predictions.

The mass density also has a characteristic decay length towards the bulk solution conditions. The eigenvalue for the volume fraction has imaginary eigenvalues with a real part $\kappa_{1,m}$ that governs the decay rate:
\begin{equation}
\kappa_{1,m}\ell_v = \sqrt{-\frac{1}{2}+\frac{1}{2}\sqrt{1+\frac{(1-\eta)^4}{2\eta (4-\eta)}}},
\end{equation}
and an imaginary part $\kappa_{2,m}$ that governs the oscillation period
\begin{equation}
\kappa_{2,m}\ell_v = \sqrt{\frac{1}{2}+\frac{1}{2}\sqrt{1+\frac{(1-\eta)^4}{2\eta (4-\eta)}}}
\end{equation}
For a very dilute gas, $\eta\rightarrow 0$, $\kappa_{1,m}\rightarrow \infty$, and the oscillations decay rapidly. However, in the limit of dense solutions, $\eta\rightarrow 1$, the effective decay length becomes:
\begin{align}
    \frac{1}{\kappa_{1,m}}\approx \sqrt{\frac{6}{10}}\frac{d}{(1-\eta)^2}\approx 0.77 \frac{d}{(1-\eta)^2}
\end{align}
using the definition of $\ell_v$. Even though the bulk density for liquids is considerably lower than 1, this is a reasonable approximation for the decay rate. Therefore, the mass density oscillations decay slowly as the overall filling fraction goes closer to 1. Also, the oscillation period goes to:
\begin{align}
    \frac{2 \pi}{\kappa_{2,m}}=\frac{2\pi d}{\sqrt{40}}\approx d
\end{align}
so layers of one ionic diameter thickness form. Again, we stress that the coupling between the density and electrostatics could be pronounced for asymmetric ions--leading to extended screening lengths. Furthermore, asymmetric liquids can more efficiently pack closer to complete filling--as would be expected as ions are added to solvent at large concentrations. Even so, for nonlinear response, at high potentials, the mass and charge densities are intrinsically linked, as studied in the main text.

\section{Local density approximation formula}
The \color{black} local density approximation\color{black} formula used as a benchmark in Fig. 3 is given by Ref. [33]. It assumes an excess chemical potential for each species of:
\begin{equation}
    \mu_i^\mathrm{ex}=-\ln\left(1-\sum_i c_i/c_{max} \right)
\end{equation}
The parameter $\gamma=2 c_0/c_{max}$ arises when describing the filling in the bulk. Due to maximal packing constraints of 0.63 for a random close packed mixture of hard spheres \cite{TUE}, we can relate $\gamma$ to $\eta$. In this case, $\gamma=\eta/0.63$. The differential capacitance can be derived analytically as:
\begin{equation}
    C=\frac{\epsilon}{\lambda_D}\cdot \frac{\cosh \left(\frac{u_{0}}{2}\right)}{1+2 \gamma \sinh ^{2}\left(\frac{u_{0}}{2}\right)} \cdot \sqrt{\frac{2 \gamma \sinh ^{2}\left(\frac{u_{0}}{2}\right)}{\ln \left[1+2 \gamma \sinh ^{2}\left(\frac{u_{0}}{2}\right)\right]}}
\end{equation}
where $u_0$ is the dimensionless surface potential, $\phi_0 \beta e$.
The charge stored in the layer via this approximation is given by:
\begin{equation}
    q_{s}=-\operatorname{sgn}(u_0)2e\lambda_D c_0 \sqrt{\frac{2}{\gamma}} \sqrt{\ln \left(1+2 \gamma \sinh ^{2}\left(\frac{u_0}{2}\right)\right)}.
\end{equation}

\section{Comparison to simulation data set}
In the main text, only two sample data sets were presented. Here, in Fig. S4, we show the concentration profiles for a wider set of charge densities. We find that the theory has similar features, but is not oscillatory enough in general.  In Fig. S5, we also provide an analogous set of plots to Fig. 2(b), but for the simulations. We find that there is a qualitative and quantitative match for the overscreening to overcrowding regimes in the simulation and in the theory.

\begin{figure*}[h]
\begin{minipage}[h]{0.4\linewidth}
\centering
\includegraphics[width=0.9 \linewidth]{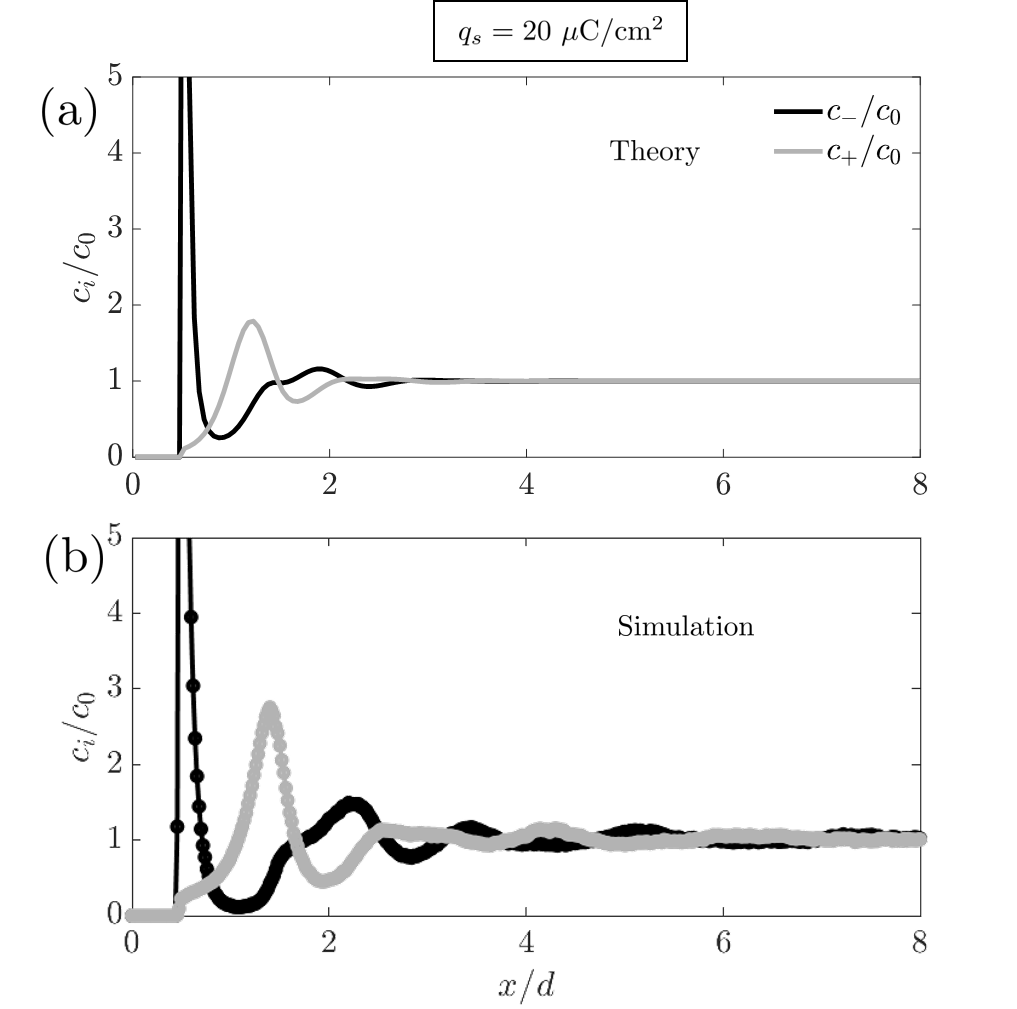}
\end{minipage}
\begin{minipage}[h]{0.4\linewidth}
\centering
\includegraphics[width=0.9 \linewidth]{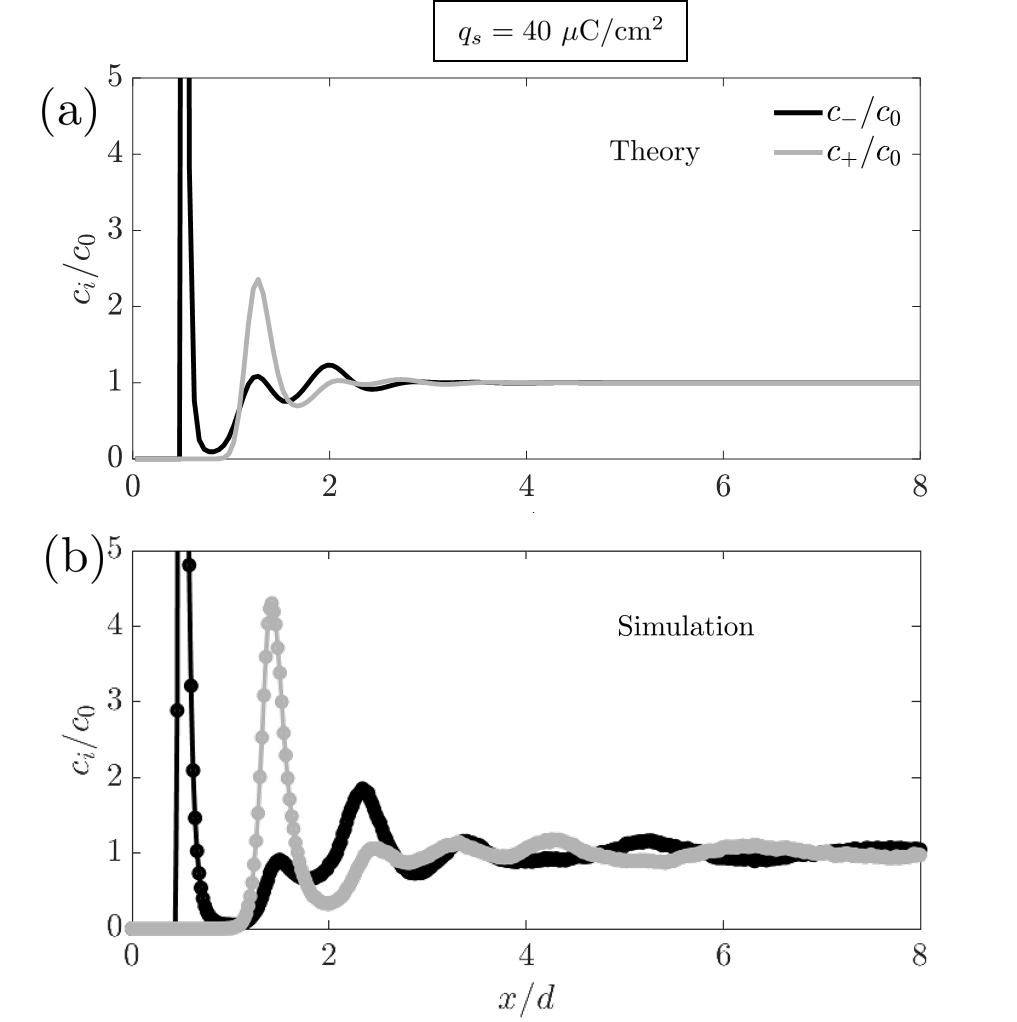}
\end{minipage}
\begin{minipage}[h]{0.4\linewidth}
\centering
\includegraphics[width=0.9 \linewidth]{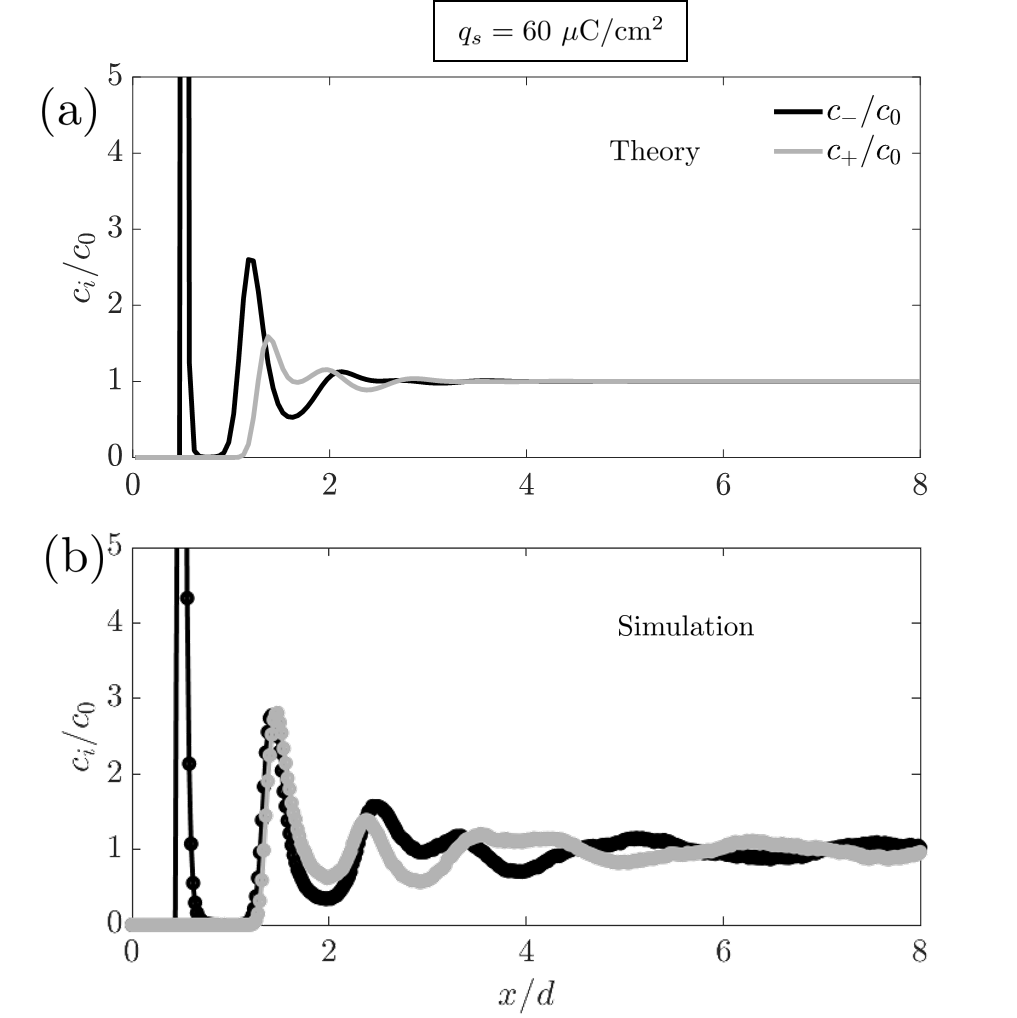}
\end{minipage}
\begin{minipage}[h]{0.4\linewidth}
\centering
\includegraphics[width=0.9 \linewidth]{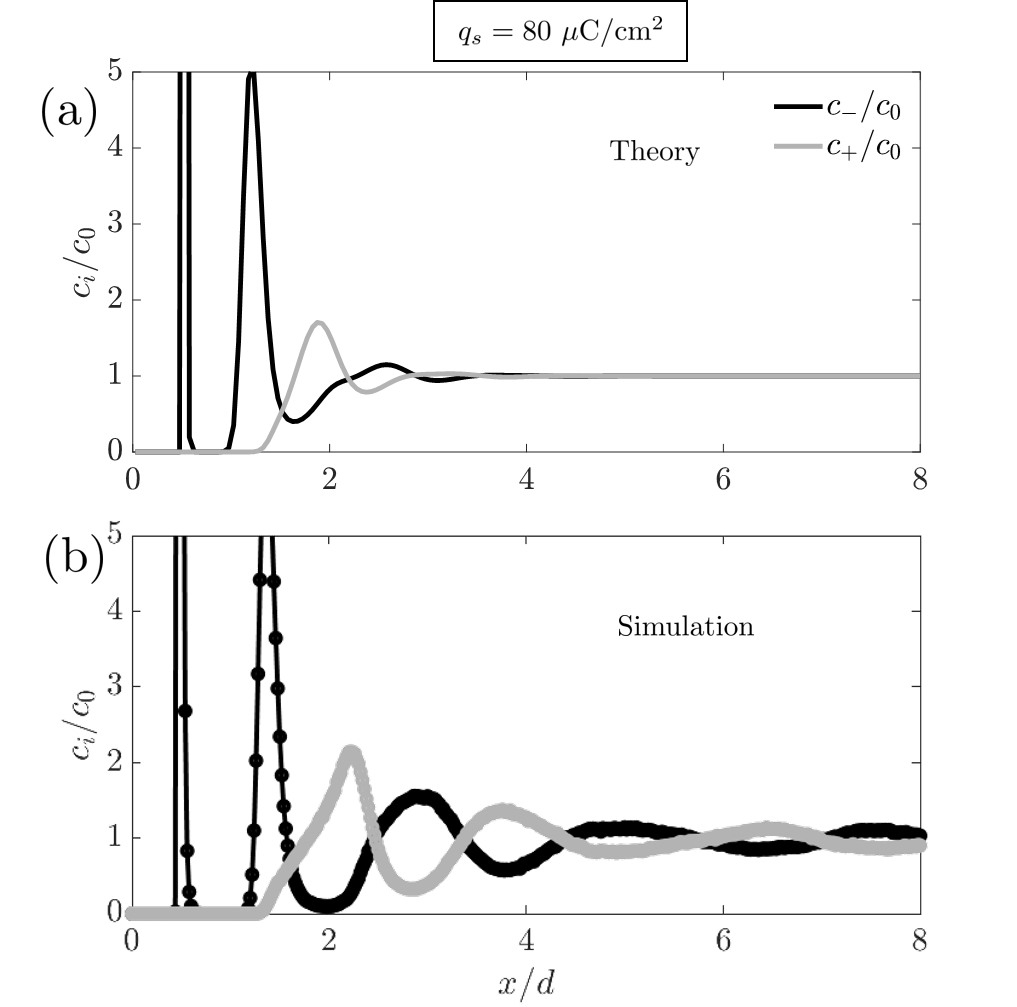}
\end{minipage}
\begin{minipage}[h]{0.4\linewidth}
\centering
\includegraphics[width=0.9 \linewidth]{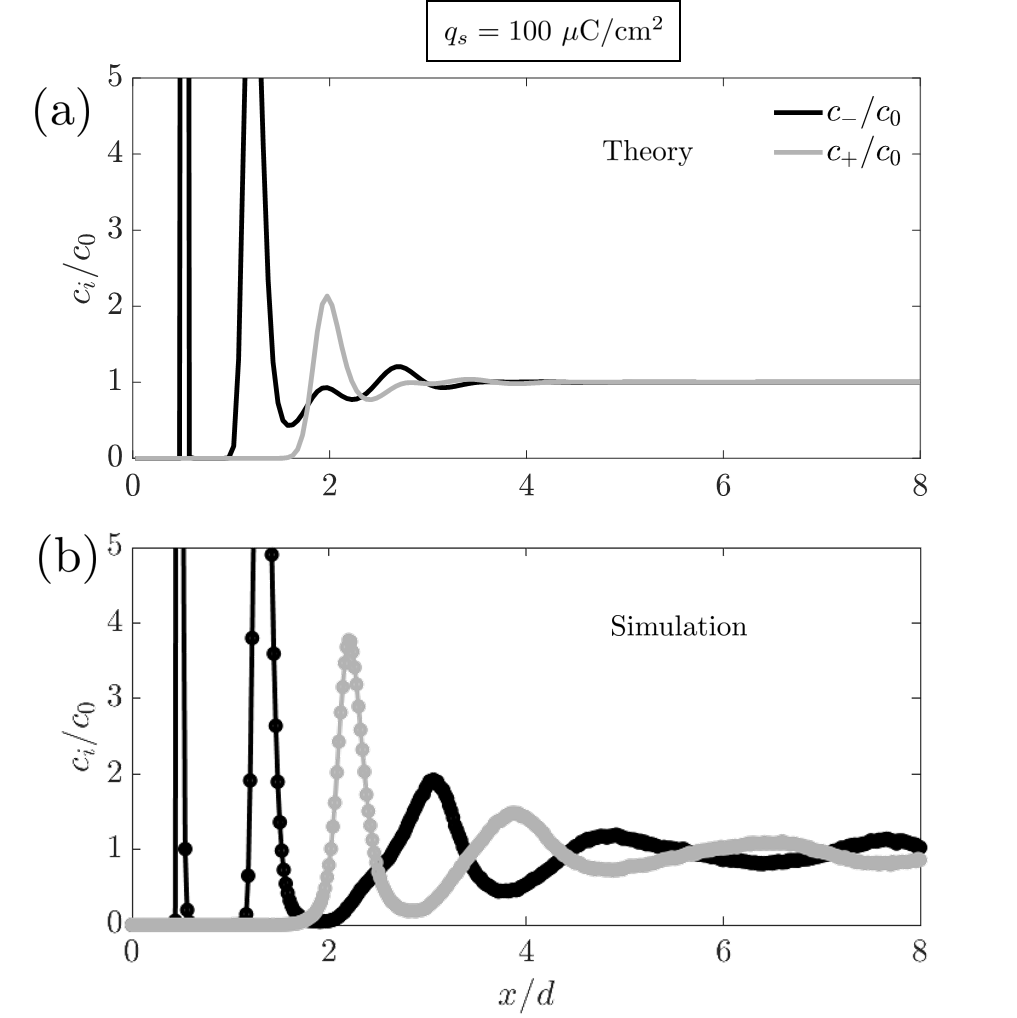}
\end{minipage}
\begin{minipage}[h]{0.4\linewidth}
\centering
\includegraphics[width=0.9 \linewidth]{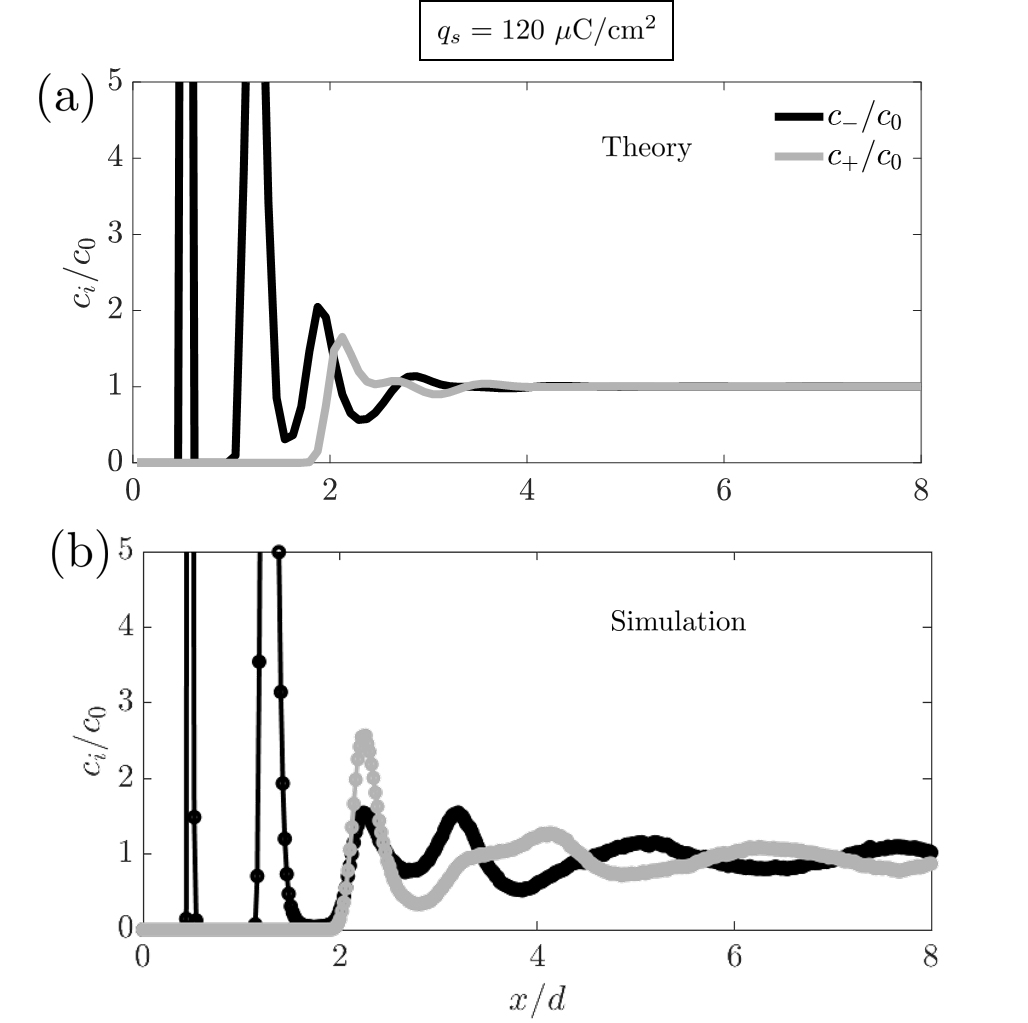}
\end{minipage}
\caption{Comparison of theory (a) and simulation (b) concentration profiles for a variety of charge densities. The MD simulation of the representative ionic liquid otherwise has the same parameters as in Figure 4.  }
\end{figure*}

\begin{figure}[b!]
\centering
\includegraphics[width=0.9 \linewidth]{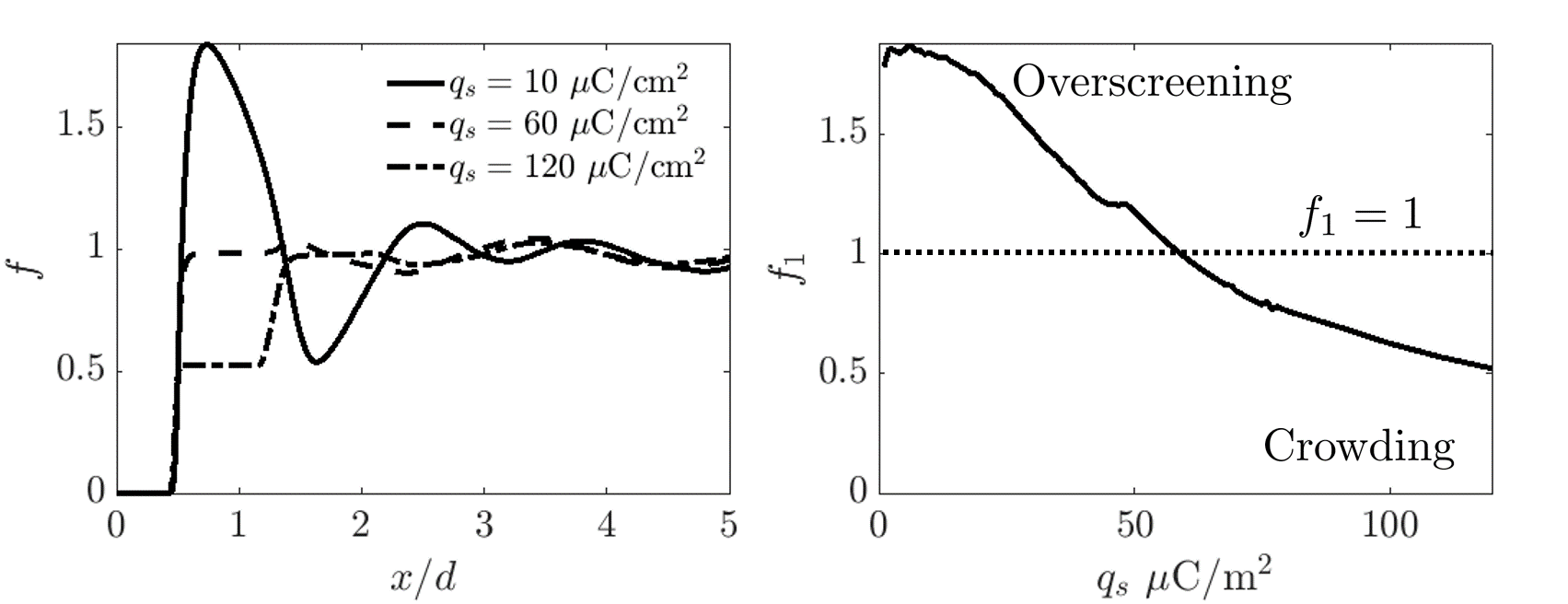}
\caption{The overscreening to overcrowding transition in the MD simulations of the representative ionic liquid. (a) The cumulative fraction of screening charge as a function of distance from the surface for different charge densities. (b) The maximum cumulative fraction of screening charge in the first layer, $f_1$ as a function of surface charge density.}
\end{figure}
\newpage

\section{Numerical implementation}

We seek a numerical solution of the system of Eqs.~(2),~(4),~(5) and (6). In Eq.~(2), we show our modified Poisson equation - the main result of this work - where the \textit{weighted} charge density appears. The weighted charge density is related to the concentrations of ions, which are in turn related to the \textit{weighted} electrostatic potential and \textit{weighted} excess chemical potential as outlined in Eq.~(6). The excess chemical potential from the finite size of ions is obtained from minimising Eq.~(4) with respect to the weighted ion concentrations. In Eq.~(6) it is shown how these weighted concentrations are related to the local concentrations; that being, a convolution with a weighting function. 

In the limit of the radius of an ion, $R$, tending to zero, the weighting functions become delta functions, and the modified Poisson-Boltzmann equation is recovered (modified because of the presence of the Carnahan-Starling excess chemical potential). It is first instructive for one to solve this modified Poisson-Boltzmann equation using a numerical finite difference solver.

To be able to solve the integro-differential equation numerically, one needs to introduce numerical forms for the convolutions. There are four of these that need to be implemented. We will outline here how to solve the integro-differential equation numerically in 1D. In 1D, the weighting function formulas must be modified~\cite{roth2010fundamental}. They become:
\begin{align}
    &w_v(x-x^\prime)=\frac{\pi\left(R^2-(x-x^\prime)^2\right)}{v}\Theta\left(R-\mid x-x^\prime\mid\right)\\
    &w_s(x-x^\prime)=\frac{1}{2 R}\Theta\left(R-\mid x-x^\prime\mid\right).
\end{align}

We construct a uniform grid with cell centers at $x_i$ and spacing $\Delta x=x_{i+1}-x_i$. The first cell center, $x_{1}$, is placed at $x=\Delta x/2$. For ease of performing convolution integrals, we choose a grid in which an integer number of $\Delta x$ equals the ionic radius. We found $\Delta x = R/10$ converges the numerical implementation here. We calculate the electric flux on the boundaries of cells as:
\begin{equation}
\begin{split}
    &D_{i+\frac{1}{2}}=\epsilon \frac{\phi_i-\phi_{i+1}}{\Delta x}\\
    &D_{i-\frac{1}{2}}=\epsilon \frac{\phi_{i-1}-\phi_{i}}{\Delta x}
\end{split}
\end{equation}

We then enforce Eq. (2) at each cell center.
\begin{equation}
    D_{i+\frac{1}{2}}-D_{i-\frac{1}{2}}=\Delta x \bar{\rho}_{e,i}
\end{equation}


One needs to introduce a numerical form for the weighted concentration of each species. The convolution at each cell center is calculated via the following numerical integration:
\begin{equation}
    \bar{\rho}_i=\frac{\sum_j w_{ij} \rho_j \Delta x}{\sum_j w_{ij} \Delta x}.
\end{equation}
where $w_{ij}$ corresponds to evaluating the weighting function at $w(x_i-x_j)$. Note that since many of the entries for $w_{ij}$ are zero, we can simplify the numerical convolution as:
\begin{align}
    \bar{\rho}_i=\frac{\frac{\Delta x}{2}\rho_{i-n}w_{i,{i-n}}+\sum_{j=i-n+1}^{j=i+n-1} w_{ij} \rho_j \Delta x+\frac{\Delta x}{2}\rho_{i+n}w_{i,{i+n}}}{\frac{\Delta x}{2}w_{i,{i-n}}+\sum_{j=i-n+1}^{j=i+n-1} w_{ij} \Delta x+\frac{\Delta x}{2}w_{i,{i+n}}}.
\end{align}
where $n$ is the number of cells in 1 ionic radius (10 in this case). Note that the first and last term in the numerator are divided by 2 due to the edge of the weighting function occurring at those cell centers. We perform the same numerical operation to find the weighted electrostatic potential and weighted excess chemical potential at each cell center.

The boundary conditions at $x=0$ is applied by setting a fictitious cell at $x=-\Delta x/2$:
\begin{equation}
    \phi_0=\phi_1+\frac{q_s}{\epsilon \Delta x}
\end{equation}
and we apply the boundary condition for the bulk at $x=L$ (sufficiently far from 0, here $L=60R$):
\begin{equation}
    \phi_{N+1}=\phi_N.
\end{equation}

The steps to solve the equations are as follows:
\begin{enumerate}
    \item We guess a $\{\phi_i\}$ and $\{\mu^\mathrm{ex}_i\}$ at each grid point. A guess can be obtained from Poisson-Boltzmann or solving the modified Poisson-Boltzmann mentioned above.
    \item We calculate the convolutions of $\phi$ and $\mu^\mathrm{ex}$ to give the cell centered weighted electric potential, $\{\bar{\phi}_i\}$ and cell centered weighted excess chemical potential, $\{\bar{\mu}^\mathrm{ex}_i\}$. For $x<R$, the values of $\{\bar{\phi}_i\}$ and $\{\bar{\mu}^\mathrm{ex}_i\}$ are arbitrary, since the ionic densities in this region are zero.
    \item We compute the local ionic densites at each cell center from Eq.~(6). We assume that the ionic densities are zero for all $x<R$ due to hard sphere interactions with the flat surface.
    \item We compute the weighted filling fraction, $\bar{p}$, and weighted charge density, $\bar{\rho}_e$, from the local ionic densities.
    \item  At every cell center, we compute the residuals from Eq.~(2) [represented numerically in Eq.~(S30)] and the residuals of the relationship between the excess chemical potential and $\bar{p}$: $\beta\mu^\mathrm{ex}_i= (8\bar{p}_i - 9\bar{p}_i^{2} + 3\bar{p}_i^{3})/(1 - \bar{p}_i)^{3}$.
    \item We iterate on the choices of $\{\phi_i\}$ and $\{\mu^\mathrm{ex}_i\}$  until the residuals in the above step are zero.
\end{enumerate}

These are the key parts to be able to solve the integro-differential equation numerically, which is not significantly more complicated than solving the modified Poisson-Boltzmann equation in the limit of $R\rightarrow0$. Therefore, we believe that this formulation of the problem permits itself to be extended to more complicated problems, as outlined in the main text.

\newpage
\section{Molecular Dynamics Simulations Details} 
All Molecular Dynamics Simulations are performed using the LAMMPS simulation package~\cite{plimpton1995parallel}. The NVT ensemble is used with a Nose-Hoover thermostat maintaining the temperature at 300~K. The cell consists of two bounding surfaces enclosing a box that is 3~nm$\times$3~nm$\times$10~nm. The first two dimensions are periodic, while the third (normal to the surfaces) is not. Surface atoms are placed on an fcc lattice (1 0 0), and have a diameter of 0.5~nm. The surface charge density is distributed evenly between all the atoms on the surface, and simulations are performed with surface charge $q_s=$1-120 $\mu$C/cm$^2$ in 1 $\mu$C/cm$^2$ increments. The anions and cations are assumed to be univalent with 0.5~nm diameter. The background dielectric constant is chosen to be a constant of 10. The concentration profiles are generated by time averaging the concentration within 0.01~nm bins. The simulations are initialized with 271 anions and 271 cations with random configurations given by the open-source softwares PACKMOL~\cite{martinez2009packmol} and FFTOOL~\cite{padua2015fftool}. The system is initialized for 5~ns, and then the production run is another 5~ns of simulation time (with 1~fs time steps). Long range electrostatics are calculated using the Particle-Particle Particle Mesh (PPPM) method beyond a cutoff of 1.2~nm. All ions have electrostatic interactions and hard sphere interactions. The hard-sphere interactions are approximated using Lennard-Jones interaction with a cutoff at the ion diameter, for a purely repulsive force ($\sigma=0.5$ nm, $\varepsilon=18.5 k_B T$). The  The atoms are visualized in Fig.~1 using the open-source Visual Molecular Dynamics (VMD)~\cite{humphrey1996vmd}.
\bibliography{REF.bib}